\documentclass[10pt,journal,compsoc]{IEEEtran}

\ifCLASSOPTIONcompsoc
   \usepackage[nocompress]{cite}
\else
    \usepackage{cite}
\fi

\ifCLASSINFOpdf
 
\else
 
\fi

\usepackage{graphicx}
\usepackage{xcolor}
\usepackage{tabularx}

\usepackage{authblk}

\begin{document}

\title{Challenges of Implementing Agile Processes in Remote-First Companies}

% \author{Lulit Asfaw}
% \author{Mikael Clemmons}
% \author{Cody Hayes}
% \author{Elise Letnaunchyn}
% \affil{College of Computing and Software Engineering,\\ Kennesaw State University, GA, USA
%  \\{lasfaw, mclemmo2, chayes86, eletnaun}@students.kennesaw.edu
% }

\author{\IEEEauthorblockN{Lulit Asfaw\IEEEauthorrefmark{1}, Mikael Clemmons\IEEEauthorrefmark{1}, Cody Hayes\IEEEauthorrefmark{1}, Elise Letnaunchyn \IEEEauthorrefmark{1},
Elnaz Rabieinejad\IEEEauthorrefmark{2}}
	\IEEEauthorblockA{
	 %	\IEEEauthorrefmark{1}Computer Science and Engineering Department, Indian Institute of Technology Guwahati, Guwahati, India, email: zolfaghari.b1975@gmail.com \\
	 \\
		\IEEEauthorrefmark{1}College of Computing and Software Engineering,\\ Kennesaw State University, Marietta, GA USA
 \\lasfaw@students.kennesaw.edu, mclemmo2@students.kennesaw.edu, chayes86@students.kennesaw.edu,
 eletnaun@students.kennesaw.edu,
 \\
					\IEEEauthorrefmark{2}Cyber Science Lab, School of Computer Science, University of Guelph,	Ontario, Canada \\  erabiein@uoguelph.ca  \\
							}}

%\author{Michael~Shell,~\IEEEmembership{Member,~IEEE,}
        %John~Doe,~\IEEEmembership{Fellow,~OSA,}
        %and~Jane~Doe,~\IEEEmembership{Life~Fellow,~IEEE}% <-this % stops a space
%\IEEEcompsocitemizethanks{\IEEEcompsocthanksitem M. Shell was with the Department
%of Electrical and Computer Engineering, Georgia Institute of Technology, Atlanta,
%GA, 30332.\protect\\

%E-mail: see http://www.michaelshell.org/contact.html
%\IEEEcompsocthanksitem J. Doe and J. Doe are with Anonymous University.}% <-this % stops an unwanted space

%{Shell \MakeLowercase{\textit{et al.}}: Bare Demo of IEEEtran.cls for Computer Society Journals}

\IEEEtitleabstractindextext{%
\begin{abstract}
\textcolor{black}{The trend of remote work, especially in the IT sector, has been on the rise in recent years, and its popularity has especially increased since the COVID-19 pandemic. In addition to adopting remote work, companies also have been migrating toward managing their projects using agile processes. Agile processes promote small and continuous feedback loops powered by effective communication. In this survey, we look to discover the challenges of implementing these processes in a remote setting, specifically focusing on the impact on communication. We examine the role communication plays in an agile setting and look for ways to mitigate the risk remote environments impose on it. Lastly, we present other miscellaneous challenges companies could experience that still carry dangers but are less impactful overall to agile implementation.}
\end{abstract}

}

\maketitle

\IEEEdisplaynontitleabstractindextext

\IEEEpeerreviewmaketitle

\section{Introduction}

Agile, in the context of software development, is a term most often related to project management methodologies that have been cultivated and refined over many years. The earliest inspirations for agile date back to 1948 and 'The Toyota Way, but it was not until 2001 that the most commonly accepted principles in agile today were documented in "The Manifesto for Software Development" \cite{1}. The manifesto outlined four values and twelve supporting principles, and it is one of those twelve principles that are of interest to us when analyzing agile methods in remote work \cite{a1,a2}. The important principle states, "The most efficient and effective method of conveying information to and within a development team is face-to-face communication" \cite{2}. Agile methods look to promote progress through small iterations and continuous feedback loops, and any lapse of communication can delay these feedback loops. A transition to remote work has been a trend, especially within software development, where many collaborative tools allow development teams to conduct their work from anywhere. 
In this paper, we first clarify some terminology concerning agile and remote work, followed by a look into the history of the transition to working remotely. After that, we examine the agile processes that require effective communication and then presented why face-to-face communication is the best communication form to use. Then, we discuss potential solutions for improving the quality of communication and ended with other miscellaneous challenges remote work presents to an agile development process. In addition, the following section has been dedicated to the hybrid working role in decreasing remote working problems, and finally, we conclude the paper in the last section.

\section{TERMINOLOGY }\label{ExSurv}

\subsection{Key Terms}

The term "agile" is often used in conjunction with other terms like processes, methodology, project management, etc. In this paper, when mentioned alone or with different keywords, we are referring to the style of project management established in 2001 by seventeen individuals and the twelve principles they established, collectively referred to as "The Agile Manifesto." The twelve principles are referenced often within this paper and therefore are listed in Section B below. It's hard to express a singular meaning for the term "agile," but as Rick Freedman explains it, "Agile is an evolutionary and revolutionary way of thinking for enterprises" \cite{3,6986022,JAVD2015295}. Therefore, a way to think about the meaning of the word is that it is a mindset, a collection of principles, and a set of processes used to manage projects. Agile is often strongly associated with the IT sector, but it has applications outside the tech industry. The reasoning for this association is due to the backgrounds of its founders \cite{4}. 
The term "remote" is used for someone working from somewhere other than a company office. This term's main idea is that remote teams are not co-located in a physical area and, therefore, cannot engage in face-to-face communication  \cite{a3}. Companies have different approaches to remote work policies, and thus several terms to describe the nature of these approaches have been used. In the context of this paper, when referencing a remote or remote-first company, we are referring to companies that operate in a one hundred percent remote environment. Other terms you may hear are hybrid or remote-friendly, and they often refer to a partial remote/partial in-office policy \cite{5}. Hybrid and/or remote-friendly work policies are outside the scope of this survey.

\subsection{Twelve Principles of Agile Software}

\begin{itemize}
	\item {Our highest priority is to satisfy the customer through early and continuous delivery of valuable software. }
	\item {	Welcome changing requirements, even late in development. Agile processes harness change for the customer's competitive advantage.}
	\item{Deliver working software frequently, from a couple of weeks to a couple of months, with a preference to the shorter timescale.} 
		\item{Business people and developers must work together daily throughout the project.} 
		
				\item{Build projects around motivated individuals. Give them the environment and support they need, and trust them to get the job done.} 
						\item{The most efficient and effective method of conveying information to and within a development team is face-to-face conversation.} 
								\item{Working software is the primary measure of progress.
} 
										\item{Agile processes promote sustainable development.} 
												\item{The sponsors, developers, and users should be able to maintain a constant pace indefinitely.} 
														\item{Continuous attention to technical excellence and good design enhances agility.} 
																\item{Simplicity--the art of maximizing the amount of work not done--is essential.} 
																
																\item{The best architectures, requirements, and designs emerge from self-organizing teams.} 
																
																\item{At regular intervals, the team reflects on how to become more effective, then tunes and adjusts its behavior accordingly} 
\end{itemize}

\section{	HISTORY OF REMOTE WORK}
Some of the very first work-from-home gestures began around 1970. This was mainly due to inflation in the oil embargo industry. Because prices rose and fuel became so expensive, the cost was not feasible; at this time, remote work made a minimal but groundbreaking introduction to office work. In 1997 Google launched the powerful Google Search Engine, which is used globally today. A few years later, project management tools were developed and released to help team members communicate effectively. Next, the most significant technological transformation was the widespread adoption and access to broadband internet. This was historically the second largest stride towards remote work due to the introduction of programs and engines to make it more feasible \cite{6}. 
 Although these strides might not be consistently recognized as the initial steps to create the possibility of remote work, without the invention and implementation of the previously mentioned inventions, the operations of remote-first development teams would be very different. The consistent and speedy development of innovative and groundbreaking technology made many in information and technology-based roles realize that the tools they were working on allowed for more workplace flexibility and remote work \cite{a4, 7}. 
Fast forward about fifteen years of a slow increase in remote work, and the globe experiences the largest modern-day pandemic, Sars-CoV-2 [2019]. The effect that Sars-CoV-2 had globally altered relationships, workplaces, the economy, mental health, and farming. Although those are just a few things the pandemic has changed, the world of remote work became a reality in a more consistent capacity. It is fair to state that remote work became popular directly because of the pandemic and social distancing measures; however, many more factors like technology innovation and economic decline played a vital part in the rapid growth in adoption. With the combination of technology innovation and social distancing mandates critical for mitigating the healthcare system, remote work was needed and mandatory in many companies like Twitter, Google, and others\cite{8}. 
Although 2022 still has not seen an official end of the Sars-CoV-2 virus life, remote work has been combined with office work to create a friendly environment to ensure success. The impact of all the events dating back to 1970, especially the recent global pandemic, has revolutionized remote working, most prominently among technology and information workers.
\section{	AGILE AND COMMUNICATION}
Agile software development emphasizes the importance of communication between the software development team and the project stakeholders as well as within the project team \cite{2,a5}. According to a research study that aimed to explore the challenges faced by software development teams, many difficulties in software engineering are claimed to be solved by agile software development approaches \cite{9}. Some challenges include extended development periods, more considerable costs than expected, and unfulfilled project requirements.
Members of software development teams should ideally have rich interactions and have the opportunity to collaborate and have frequent face-to-face meetings. According to the study, employees in different locations are less likely to see themselves as part of the same team. That face-to-face interaction is vital for successful team collaboration \cite{9}. Face-to-face interactions are intended to help people get to know one another and build social networks that can foster trust, respect, and commitment and facilitate long-term development work across many geographic locations.
With today's technical advancements allowing seamless remote meetings, it isn't easy to understand why face-to-face communication remains so popular. Because not all team members or stakeholders may be present in the same place simultaneously, electronic communication tools are frequently used in agile projects. However, face-to-face communication provides a certain level of clarity that cannot be achieved through online communication. Face-to-face communication facilitates quick feedback loops and enhances nonverbal communication when employees interact with one another. It strengthens bonds, fosters mutual trust, and establishes centers of influence within a company \cite{1}. 
Face-to-face interaction is especially preferred in collaborative work, where more critical tasks such as daily stand-up and discovery sessions are considered. A daily stand-up is an essential aspect of the agile process since it improves team communication. Integrating good communication with face-to-face interaction as part of a team's culture aids in developing a more Agile workplace \cite{1}. The team breaks down the project idea into tangible requirements in a discovery session. Any miscommunication or misunderstanding during this process can have a devastating impact. Face-to-face communication takes less time and effort and has several dimensions via which meaning can be better conveyed, making it an ideal choice for most Agile processes.

Face-to-face communication plays a significant role in facilitating collaboration among team members from different cultural backgrounds. Teams are made up of people from different cultures and backgrounds. Cultural differences are key for a successful team because each team member can approach the problem from a different perspective and offer suggestions based on their unique background. However, cultural differences can also lead to conflicts and, consequently, project failures \cite{1}. According to Hofstede's cultural dimensions theory, cross-cultural communication is prone to misinterpretation because people from different cultural backgrounds could understand identical verbal and nonverbal messages differently \cite{10}. Face-to-face communication fosters cultural understanding by creating a trusting environment. It is thus less likely for messages to be misinterpreted. Communication is already a challenge in cross-cultural teams, and using online communication tools that do not accurately display non-verbal cues and body language can further accentuate communication issues. 
Cultural differences can be a barrier to communication, affecting one's capacity to connect with others and drive them. The Six Dimensions of Culture, developed by Hofstede, can be used to understand and work effectively with people from other cultures. It can also be used to assemble a high-performing team from individuals with diverse cultural and geographic backgrounds. Hofstede's Six Dimensions of Culture include Power Distance Index, Individualism vs. Collectivism, Masculinity vs. Femininity, Uncertainty Avoidance Index, Long- Versus Short-Term Orientation, and Indulgence vs. Restraint as seen in "Fig. 1" \cite{10}. Culture clashes can severely affect the work environment and workflow. It is essential to understand the cultural values of each team member to understand the importance of face-to-face communication further.

   \begin{figure}
    \centering
    \includegraphics[width=8cm]{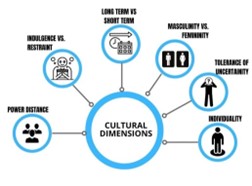}
    \caption{	Hofstede’s Six Dimensions of Culture}
    \label{fig:life}
\end{figure}

\section{	HOW TO IMPROVE COMMUNICATION QUALITY}
Focusing on improved communication quality through a communication plan begins with understanding what asynchronous and synchronous forms of communication are and how each can be utilized to benefit the remote teams \cite{11}. Synchronous communications are when team members communicate with each other in real-time. Examples of this communication include conference calls, instant messaging, or Teams/Zoom meetings. Face-to-face communications in the office, even if the conversation is in passing, would also be synchronous.   On the other hand, asynchronous communications allow team members to respond when each member’s schedule permits, such as emails, group chats, or even voice mails and messages.  
Although synchronous communication is crucial in many agile processes, whenever it is not required, asynchronous communication is the best way for team members and other stakeholders within remote teams to interact \cite{11}. This approach provides flexibility for team members across different time zones, allows for team members to stay focused on their tasks at hand, reduces the stress of responding to work communications outside of their working hours, and also allows for team members to think through problems and ideas, in turn reducing impulse responses.
Furthermore, without face-to-face communication, team members can miss contextual clues given by body language and other nonverbal, visible cues, so keeping messages as clear and to the point as possible is good practice also. Even incorporating the use of emojis can give more precise context. For employers, avoiding micromanaging is imperative for remote teams to succeed, which is one of the main reasons Agile is so effective. Team members feel more responsible and motivated to perform their work without worrying about micromanaging; think back to the Agile principle about trusting the employees to do their job \cite{a6}. Having the right tools in place, such as project management software, video calling, or even a virtual “break room” where team members can make small talk, all help to emulate better being in an office and help team members feel more connected and motivated \cite{12}.
\section{	OTHER CHALLENGES IN AGILE/REMOTE WORK}
Being remote can also directly challenge or strain Agile’s core values. As a core value, “Individuals and Interactions over processes and tools” emphasizes the importance of direct collaboration, especially in those activities best suited for face-to-face interaction \cite{2}. While having processes and tools in place provides teams the ability to be Agile, the focus is still on having motivated people working together effectively as a team to deliver value to the customer. However, in a remote team, processes and tools play a more significant role in allowing the team to collaborate and effectively provide the same expected value.  
Additionally, remote first companies and their teams may find difficulty in adhering to the Agile principle of “Build projects around motivated individuals. Give them the environment and support the need and trust them to get the job done.” \cite{1}.  This applies more to the company because motivated teams will produce more, higher quality work\cite{a7}. If the employer treats employees like drones who cannot be trusted to perform their jobs, employees will eventually act like drones who cannot be trusted to perform their jobs. And it’s here where the employer and the remote teams can struggle to find the balance to allow the team to stay motivated and be successful, as monitoring activity and over-communicating to keep remote teams on task can lead to micromanaging and subsequently demotivating the teams to perform.  
With over-communicating having potential downstream negative impacts on the remote teams as the aforementioned more desirable face-to-face communications are not able to be had for Agile processes that typically require such, having an effective communication plan for the teams is now more essential. In an office environment, face-to-face communication is much more accessible and likely to happen naturally outside of scheduled meetings just by proximity \cite{12}. In remote teams, team members may not even be in the same time zones, so even more typical communication methods such as phone calls or video chats can be challenging to coordinate. Also, with many remote team members working from home offices, without a clear delineation of the work environment from the home environment, communications can become less professional and more casual, blurring the work-life boundaries, which can lead to other stressors. Moreover, getting back to the tools utilized by remote teams, team members may have technical limitations due to poor internet connectivity and availability, in addition to the equipment they may use for work. And some forms of communication can even lead to friction amongst team members due to misinterpreting context in emails or other messages.

\section{Hybrid working; one step ahead of the remote working}
In the post-pandemic age and with revolution, It is doubtful that companies will return to an utterly face-to-face working method. Remote working has many advantages for both employees and employers \cite{13}. For instance, employers can hire experts regardless of geographical boundaries. Also, employees can work without spending extra energy and money on daily commutes. However, as mentioned earlier, remote working challenges and face-to-face communication's importance for improving performance is not neglectable.

Hybrid working can be a critical key and allow companies to use both remote and face-to-face working advantages. Hybrid working gives employees the right to work in the office or remotely \cite{13}. In fact, this method tries to satisfy different employees' preferences. Furthermore, some employees prefer to collaborate with team members in-person or, due to some technical difficulties, don't have enough performance remotely, can work in-office. Although Hybrid working can decrease remote working problems, it has some challenges that must be considered .  As long as the working space has been split into office and virtual, it is essential that both groups feel equal \cite{14}. In other words, remote or office employees should have equal rights in participation, representing their ideas, etc. In this way, IT leaders should provide some facilities for remote employees to feel comfortable such as rearranging chairs and tables around cameras and microphones during meetings, improving sound systems, providing the necessary equipment for remote employees, and setting some in-person meetings for all employees to talk about their problems \cite{15}.If a company could make these minor changes can turn their company into a flexible and effective environment that can use Agile methodology in the best way.

\section{	CONCLUSION}
	In conclusion, a combination of Agile project management and remote work has become more popular, especially in the IT sector, but for different reasons. Agile solves many development problems introduced by the traditional waterfall model but is highly dependent on effective communication. Remote work has been made possible by advancements in technology that allow collaboration amongst project teams located worldwide. Still, its adoption has been influenced in recent years due to the COVID-19 pandemic \cite{SHAHID2021103751}. This pandemic has led more companies to adopt "remote-first" policies that allow, and even further, expect their employees to work from somewhere other than a company office. In co-located teams, communication is more accessible and often happens naturally due to the proximity of team members. In remote teams, new risks are introduced in remote environments and, if not handled properly, can disrupt previously successful Agile implementations.
	As previously mentioned, effective communication is essential to the success of Agile implementations. As stated in the guiding principles of "The Agile Manifesto," face-to-face communication is the most effective form of communication. Agile's success depends on short, continuous feedback loops that deliver feedback to development teams and allow for incremental development cycles. Therefore, poor communication can impose delays that interrupt these crucial feedback loops. In remote-first companies, where accurate face-to-face communication is impossible, knowledge of the risks to communication is key in mitigating them. Many agile processes greatly benefit from in-person communication, where body language and other visual cues are supplemental methods of conveying ideas. Cultural differences can significantly impact communication, and remote teams naturally promote more culturally diverse teams. Face-to-face communication lessens misinterpretations of cultural differences, which again is not possible in remote teams.
	Lastly, there are many things remote-first companies can do to promote effective communication within remote teams. Still, even with that, they may face other potential issues in remote agile teams. Other fundamental principles of Agile state that success depends on motivated team members, and lack of human interaction can often cause mental distress. It is crucial to the success of Agile that companies provide an environment that fosters communication and encourages its employees. With the correct measures, Agile can still be successful in a new, remote-first world.

\bibliographystyle{IEEEtran}
\bibliography{References}

\end{document}